\documentstyle[12pt]{article}
\newcommand{\sect}[1]{\setcounter{equation}{0}\section{#1}\indent}
\renewcommand{\theequation}{\thesection.\arabic{equation}}
\renewcommand{\thefootnote}{\fnsymbol{footnote}}
\newcommand{\EQ}{\begin{equation}}
\newcommand{\EN}{\end{equation}}
\newcommand{\bea}{\begin{eqnarray}}
\newcommand{\ena}{\end{eqnarray}}
\newcommand{\vs}[1]{\vspace{#1 mm}}

\renewcommand{\a}{\alpha}
\renewcommand{\b}{\beta}
\renewcommand{\c}{\gamma}

\newcommand{\e}{\epsilon}

\newcommand{\G}{\Gamma}

\newcommand{\uda}{\nearrow \kern-1em \searrow}
\newcommand{\half}{{1 \over2}}
\newcommand{\NP}[1]{Nucl.\ Phys.\ {\bf #1}}
\newcommand{\PL}[1]{Phys.\ Lett.\ {\bf #1}}
\newcommand{\CMP}[1]{Comm.\ Math.\ Phys.\ {\bf #1}}

\newcommand{\IJMP}[1]{Int.\ Jour.\ Mod.\ Phys.\ {\bf #1}}

\newcommand{\mone}{$P^{1,1,2,8,12}[24]$} 
\newcommand{\mtwo}{$P^{1,1,2,4,4}[12]$}
\newcommand{\mthree}{$P^{1,1,2,4,8}[16]$}
\newcommand{\mfour}{$P^{1,1,2,4,4,4}[8,4]$}
\newcommand{\third}{{1\over 3}}
\newcommand{\fourth}{{1\over 4}}
\makeatletter
\def\eqnarray{%
 \stepcounter{equation}%
 \let\@currentlabel=\theequation
 \global\@eqnswtrue
 \global\@eqcnt\z@
 \tabskip\@centering
 \let\\=\@eqncr
 $$\halign to \displaywidth\bgroup\@eqnsel\hskip\@centering
 $\displaystyle\tabskip\z@{##}$&\global\@eqcnt\@ne
 \hfil$\displaystyle{{}##{}}$\hfil
 &\global\@eqcnt\tw@$\displaystyle\tabskip\z@{##}$\hfil
 \tabskip\@centering&\llap{##}\tabskip\z@\cr}
\makeatother

\begin{document}

\begin{titlepage}
\setcounter{page}{0}
\begin{flushright}
EPHOU 97-001\\
January 1997\\
\end{flushright}

\vs{6}
\begin{center}
{\Large Enhanced Gauge Symmetry in Three-Moduli Models of Type-II String and Hypergeometric Series}

\vs{6}
{\large
Hisao Suzuki\footnote{e-mail address: hsuzuki@phys.hokudai.ac.jp}}\\
\vs{6}
{\em Department of Physics, \\
Hokkaido
University \\  Sapporo, Hokkaido 060 Japan} \\
\end{center}
\vs{6}

\centerline{{\bf{Abstract}}}
The conifold singularities in the type-II string are considered  as the points of phase transition. In some cases, these singularities can be understood in the framework of  the conventional fields theores as the points of  enhanced gauge symmetry. We consider a class of three moduli Type-II strings. It is shown  the periods can be written in the form of  hypergeometric series around the singular points in these models. The leading expansion around the conifold locus turns out  to be described by Appell functions. In one singular point, we observe the enhanced gauge symmetry of $SU(2)\times SU(2)$ independent of the models.  Around another conifold locus, however, the resulting expression of the Appell functions depends on the models. We  examine the result by considering a relation between these Appell functions and underlying Riemann surfaces. 
\end{titlepage}
\newpage

\renewcommand{\thefootnote}{\arabic{footnote}}
\setcounter{footnote}{0}
\sect{Introduction}
There has been amazing development in understanding of the non-perturbative aspect of the N=2 supersymmetric theories, which have occurred both in the rigid field theories and in the string theores. In the field theories, it is initiated by Seiberg and Witten\cite{SW}. In the string side, it begun in the work of Hull and Townsend\cite{HT} as the heterotic-type II duality. These considerations correlate in four dimension as the theory of duality in string theories which has much  caluculability\cite{FHSV,KV}. In the type-II side, the necessary condition for having heterotic dual is that the Calabi-Yau manifolds are fibration of $K3$ surfaces\cite{KLM,AL,Aspinwall}. Although the number of these $K3$-fibred Calabi-Yau manifolds is very large\cite{AKMS}, the number of models having defenite dual pair is very small at present.  The study of the non-perturbative aspects of these models are quite important for analyzing  the phase space of the whole moduli space of the strings, which!
 can be performed  without the e

xistence of the dual theories in the type-II theories.. 

An interesting phase transition in these theories  is the enhanced gauge symmetry.  In the type-II string, the points of the gauge enhancement appear at the conifold singularities\cite{KMKMP}. However, it turns out that the variety of such  singularities are very large. Namely, the conifold singularities not only describe physics which can be understood in terms of Yang-Mills theories but also contain theories possessing new type of phase transitions which cannot be understood by any conventional physics\cite{Lerche}. In this sense, it seems of particular interesting to understand what kind of phase transition we can observe at generic singularities of the Calabe-Yau space. At low energy limit, these considerations may generate new kind of rigid theories or the theories without weak coupling phases. 

As for the  understanding of the moduli space, the structure of the moduli has been studied quite systematically  at large moduli. Correspondingly, the periods and prepotential are mostly studied by large moduli expansions\cite{HKTY}. This is benefitted from the fact that the periods can be expressed by hypergeometric series\cite{Batyrev,HKTY} because of which we can calculate the prepotentials quite systematically. In other words,  these considerations are supported by  the universality of the structure at the large moduli limit. Contrary to these consideration, there is no known theories about expression of the periods around generic conifold locus. Therefore, it seems interesting to find models whose conifold singularities can be analyzed systematically, which may hopefully serve as a first step for constructing a systematic expansion around conifold locus for generic  Calabi-Yau manifolds. 

In this paper, we deal with  some particular class of three moduli models. Specifically, the models we consider are $P^{1,1,2,8,12}[24], P^{1,1,2,4,4}[12], P^{1,1,2,4,8}[16]$ and $P^{1,1,2,4,4,4}[8,4]$. These models are known to have an interesting structure. Namely, these are fibrations of K3 which are fibrations of the torus. The first model \mone {} is known to have defenite dual heterotic strings\cite{KV} and the non-perturbative behavior of the model has been nicely studied\cite{KKLMV,KLMVW}. The interesting observation about the underlying K3 surface of the models \mtwo{} and \mthree{} have been made in Ref.\cite{LY}. In addition, we  include  the last model \mfour{} in this paper. We will show that the periods of these models can be expressed in the form of hypergeometric series around the conifold singularities.  Although the method we will use in this paper cannot be applies for all Calabi-Yau models, we hope our resulting expression can be generalized to generic mode!
ls. 
We will also consider a low energy limit of  the expression of the periods and find that these described by Appell functions. We will consider the relation between the functions and the low-energy Riemannian surfaces. 

In the next section, we will review the structure of the models and periods at large moduli expansion. We will show how to obtain the expansion of the periods out of the expression at large moduli. 

In section 3,  we will obtain a expansion of the periods around a singular point. It will be shown that the periods can be expressed by hypergeometric series around the singularity. In the leading order, we will find that the some solutions reduce to the Appell functions of the type $F_4$, which decompose into the hypergeometric functions independent of these models. From these resulting expression, we will identify the singularity as a point for an enhanced gauge symmetry for $SU(2)\times SU(2)$.  

In section 4, we consider another singular point where the $P^{1,1,2,8,12}[24]$ model is known to have an enhanced $SU(3)$ symmetry\cite{KV}. We can obtain the expansion of the Periods in terms of hypergeometric series. It will turn out that the leading order expansion can be expressed  by Appell functions whose  parameter  differs by models. 

In section 5, 
we consider the underlying Riemannian surfaces corresponding to the singular point and discuss a relation to the expression of the periods obtained in the previous section.

The last section is devoted to some discusion.

\sect{The three-moduli models and sequence of fibration}
Let us consider the Type-II string compactified on the Calobi-Yau manifold with $h_{11}=3$. A typical example for the enhanced gauge symmetry is the $K3$-fibration threefold $P^{1,1,2,8,12}[24]$. In the type-IIB side, the defining polynomial is given by\cite{HKTY}
\bea
f= \half z_1^2 + {1 \over 3}z_2^3 + {1\over12}z_3^{12}&+&{1\over24}z_4^{24}+{1\over24}z_5^{24} \nonumber\\
&-&\psi_0z_1z_2z_3z_4z_5 +{1\over6}\psi_1(z_3z_4z_5)^6 + {1\over12}\psi_2(z_4z_5)^{12},
\label{eq:eq21}\ena
where we have followed by the normalization given in Ref.\cite{KKLMV,KLMVW}.

This model is known to be the fiber of a $K3$ surface of type $P^{1,1,4,6}[12]$ which is also a elliptic fibration with fiber $P^{1,2,3}[6]$.\cite{KLM,LY} 
There are three other models having this kind of property corresponding the fact that there are mainly four type of toric description of the torus. That is, the models are $P^{1,1,2,4,4}[12], P^{1,1,2,4,8}[16]$ and $P^{1,1,2,4,4,4}[8,4]$. The defining polynomials in the type-IIB side are given by\\
\noindent$P^{1,1,2,4,4}[12]$: 
\bea
f={1\over 3} z_1^3 + {1\over3}z_2^3 + {1\over6}z_3^6 &+& {1\over12}z_4^{12}+ {1\over12}z_5^{12}\nonumber\\
 &-& \psi_0z_1z_2z_3z_4z_5 + {1\over3}\psi_1(z_3z_4z_5)^3+{1\over6}\psi_2(z_4z_5)^6,
\label{eq:eq22}\ena
\noindent$P^{1,1,2,4,8}[16]$: 
\bea
f=\half z_1^2 + {1\over4}z_2^4 + {1\over8}z_3^8 &+& {1\over16}z_4^{16}+ {1\over16}z_5^{16}\nonumber\\
 &-& \psi_0z_1z_2z_3z_4z_5 + {1\over4}\psi_1(z_3z_4z_5)^4+{1\over8}\psi_2(z_4z_5)^8,
\label{eq:eq23}\ena
\noindent$P^{1,1,2,4,4,4}[8,4]$:
\bea
f_1 &=& {1\over2}z_1^2 + {1\over4}z_2^4+{1\over8}z_3^8+{1\over8}z_4^8 - z_5z_6 +\half \psi_1(z_2z_3z_4)^2 +{1\over4}\psi_2(z_3z_4)^4, \nonumber\\
f_2 &=& \half z_5^2 + \half z_6^2 - \psi_0 z_1z_2z_3z_4.
\label{eq:eq24}\ena
We list  the sequences of fibrations of the models: 
\bea
\matrix{
{\rm{Calabi-Yau}}&\rightarrow&K3&\rightarrow&{\rm torus}\cr
P^{1,1,2,8,12}[24] &\rightarrow& P^{1,1,4,6}[12] &\rightarrow& P^{1,2,3}[6],\cr
P^{1,1,2,4,4}[12] &\rightarrow& P^{1,1,2,2}[6] &\rightarrow& P^{1,1,1}[3],\cr
P^{1,1,2,4,8}[16] &\rightarrow& P^{1,1,2,4}[8] &\rightarrow& P^{1,1,2}[4],\cr
P^{1,1,2,4,4,4}[8,4] &\rightarrow& P^{1,1,2,2,2}[4,3] &\rightarrow& P^{1,1,1,1}[2,2].\cr}
\ena
There are other models described by sequence of fibrations, whose classification was made  recently in ref.\cite{KLRY}. The reason why we choose these special models is that periods can be treated in a unified way, as we are going to see below.

Following the notation of Ref.\cite{KKLMV}, we introducing the parameters by $a=\psi_0^{1\over\lambda}/\psi_1, b=\psi_2^{-2}$ and $c=\psi_2/\psi_1^2$, where $\lambda$ is $1/6, 1/3, 1/4$ and $1/2$ for \mone, \mtwo, \mthree and \mfour, respectively. 
At large moduli $a>1$, the fundamental period of the system can be obtained as
\bea
\omega_0 =
\sum_{m_1,m_2,m_3}
&{}& {\G(m_1+\lambda)\G(m_1+1-\lambda) \over \G(m_1+1)\G(m_2+1)^2\G(m_3+1)\G(m_1-2m_3+1)\G(m_3-2m_2+1)}\nonumber\\
&\times& ({1\over a})^{m_1}({b\over4})^{m_2}({c\over4})^{m_3}.
\label{eq:eq26}\ena
The other solutions of the Picard-Fuchs equation can be obtained by shifting $m_1, m_2$ and $m_3$ by $m_1+\sigma_1, m_2+\sigma_2$ and $m_3+\sigma_3$ in (\ref{eq:eq26}) and differentiating with respect to $\sigma$'s as has been written in Ref.\cite{HKTY}. For the discussion of this paper, it is better to write the fundamental period in the form:
\bea
\omega_0= &\int{ds_1\over 2\pi i}&\int{ds_2\over 2\pi i}\int{ds_3\over 2\pi i}{\G(-s_1)\G(-s_2)\G(-s_3)\G(s_1+\lambda)\G(s_1+1-\lambda) \over \G(s_1+1)\G(s_1-2s_3)\G(s_3-2s_2)}\nonumber\\
&\times&(-a)^{-s_1}(-{b\over4})^{s_2}(-{c\over4})^{s_3},
\label{eq:eq27}\ena
where the contour integrals are  taken to be circles enclosing integers.
Then the other solution can be obtained by replacing the denominator of the coefficient by $1/\G(s_2+1)\rightarrow \G(-s_2), 1/\G(s_1-2s_3) \rightarrow \G(1-s_1+2s_3)(-1)^{s_1-2s_3},$ or $1/\G(s_3-2s_2) \rightarrow \G(1-s_3+2s_2)(-1)^{s_3-2s_2}$. Note that these replacement does not change the recursion relations satisfied by the coefficients. Therefore, the replacement maps the solution to other solution of the system. In total, we have $8$ independent solutions of the Picard Fuchs equation. These form of the solutions are convenient for considering the analytic continuations which will play an important role in our discussion.

It can be shown that the structure of the discriminant is universal in the models we treat here. Therefore, we can consider the expansion around the conifold singularity in a unified way. We shall consider two type of singular points: $(a=2, b=0,c=1)$ and $(a=0,b=0, c=0)$. For \mone, the former singularity is known to be the point of gauge enhancement for $SU(2)\times SU(2)$ and the latter corresponds to the enhancement to the $SU(3)$ group\cite{KV}. 

Let us consider how we can obtain the expansions around these conifold points. Our aim is to get a systematic expansion in the form of the  hypergeometric series just like the expansion at large moduli\cite{Batyrev,HKTY}. The most difficult question to achieve the purpose is how we find ``good'' variables by which the coefficients are written in the form of hypergeometric series. At present, it is very difficult to find such  convenient variables from the Picard-Fuchs equations. Instead of just considering the equations, we here take other approach. Namely, we will attempt to  make use of the analytic continuations and quadratic transformations of the hypergeometric series\cite{HTF}. Because of the intersection of singularities of the variables, we cannot believe the validity of such analytic continuation to the region beyond the convergence region of the original expression. However, these transformations  do map the set of solution for the original variables to other set  fo!
r the transformed variables. The

refore, starting from the solutions at large moduli, we can in principle generate solutions by other variables. Note that  this does not mean that we can obtain the solutions around conifold singularities for all models by this method. This is basically due to the fact that the analytic continuation to the point of conifold is available only for the very special form of hypergeometric series\cite{Bailey}. For example, the continuation of the hypergeometric series with respect to the variable $x$ to  $x=1$ is available only for the type ${}_2F_1$ except for some special cases\cite{HTF,Bailey}. Therefore, our analysis must rely heavily on the form of the solutions. We will show in the next section that it is possible to make analytic continuation for the solutions at least for the  models constructed be sequences of fibrations.

\sect{Expansion around the point $(a=2,b=0,c=1)$}

Let us start from the analytic continuation of the fundamental period.
By shifting the indices $m_2=n_2, m_3=n_3+2n_2$ and $m_1=n_1+4n_2+2n_3$, we rewrite the fundamental period in the form:
\bea
&\omega_0&= \sum_{n_2,n_3}F(4n_2+2n_3+\lambda,4n_2+2n_3+1-\lambda;4n_2+2n_3+1;{1\over a})\nonumber\\
&\times&{\G(4n_2+2n_3+\lambda)\G(4n_2+2n_3+1-\lambda)\over \G(4n_2+2n_3+1)\G(n_2+1)^2\G(n_3+2n_2+1)\G(n_3+1)}({1\over a})^{4n_2+2n_3}({b\over4})^{n_2}({c\over4})^{n_3},
\label{eq:eq31}\ena
where $F(a,b;c;x)$ is the hypergeometric function.

In order to obtain the expansion around $a=2$, we will apply the following quadratic transformation\cite{HTF}:
\bea
F(2a,2b;a+b+\half;\half+\half z) =&{}& {\G(a+b+\half)\G(\half) \over \G(a+\half)\G(b+\half)}F(a,b;\half;z^2)\nonumber\\
&{}& + 2z{\G(a+b+\half)\G(-\half)\over\G(a)\G(b)}F(a+\half,b+\half;{3\over2};z^2).
\label{eq:eq32}\ena
From the logarithmic solution with respect to the variable $a$, we can apply a similar transformation formula.
By this change of variables we get the following two type of solutions:
\bea
&\sum_{n_1,n_2,n_3}&{\G(n_2+2n_2+n_3+{\lambda\over2})\G(n_1+2n_2+n_3+{1-\lambda\over2})\over \G(n_1+\half)\G(n_1+1)\G(n_2+1)^2\G(n_3+2n_2+1)\G(n_3+1)}\nonumber\\
&{}&\qquad\qquad\qquad ({2\over a}-1)^{2n_1}[({2\over a})^4 {b \over 4} c^2]^{n_2}[({2\over a})^2 c]^{n_3},\label{eq:eq331}\\
&\sum_{n_1,n_2,n_3}&{\G(n_2+2n_2+n_3+{\lambda\over2}+\half)\G(n_1+2n_2+n_3+{1-\lambda\over2}+\half)\over \G(n_1+{3\over2})\G(n_1+1)\G(n_2+1)^2\G(n_3+2n_2+1)\G(n_3+1)}\nonumber\\
&{}&\qquad\qquad\qquad({2\over a}-1)^{2n_1+1}[({2\over a})^4 {b \over 4} c^2]^{n_2}[({2\over a})^2 c]^{n_3}.\label{eq:eq332}
\ena

Before considering the conifold singularity, let us consider a limit $b\rightarrow 0$, which corresponds to the decompactifying limit of the fiber. In other words, the period integral reduces to the direct product of the periods for $K3$ and those for the fiber. As a result, the half of the cycles reduces to those of the $K3$ and the rest, which has logarithmic dependence with respect to $b$, suffers from the infra-red divergences caused by the decompactification. An interesting identities are known to exists\cite{KLM,LY}  in these two-moduli class of $K3$. Namely, the solutions of the Picard-Fuchs equation can be written by products of the hypergeometric functions. We are going to re-interpret their results by using the known formula of the hypergeometric series in two variables. 

For $y=0$, the solutions (\ref{eq:eq331}) and (\ref{eq:eq332}) reduce to
\bea
F_4({\lambda\over2},{1-\lambda \over 2},\half,1;({2\over a}-1)^2,({2\over a})^2 c),\nonumber\\
F_4({\lambda\over2}+\half,{1-\lambda \over 2}+\half,{3\over2},1;({2\over a}-1)^2,({2\over a})^2 c),
\label{eq:eq34}\ena
where the function $F_4$ is the Appell function of type $F_4$ which is defined as\cite{Appell,HTF}
\bea
F_4(\a,\b,\c,\c',x,y) &=& \sum_{m,n}{(\a)_{m+n}(\b)_{m+n} \over (\c)_m(\c')_n m!n!}x^my^n,\nonumber\\
(a)_n &\equiv& {\G(a+n)\over\G(a)}.
\label{eq:eq35}\ena
For these solutions, we can apply the following product identity proved by Bailey and Watson\cite{Bailey,HTF}:
\bea
&{}&F_4(\a,\c+\c'-\a-1,\c,\c';x(1-y),y(1-x))\nonumber\\
&{}&\qquad\qquad = F(\a,\c+\c'-\a-1;\c;x)F(\a,\c+\c'-\a-1;\c';y).
\label{eq:eq36}\ena
Therefore, the observation of Lian and Yau\cite{LY} that the solution of the Picard-Fuchs equations can be written as the direct product of the hypergeometric functions are related to the product identity of the Appell functions.

It is also interesting that another identity which is conjectured and proved in  the one moduli model of $K3$\cite{KV,LY}:
\bea
{}_3F_2(\a,1-\a,\half;1,1;z) = F({\a\over2},{1-\a \over2};1;z)^2,
\label{eq:eq37}\ena
can be deduced by combining the product identity \ref{eq:eq36} with the following identities:
\bea
F_4(\a,\b,\c,\c';x,x) &=& {}_4F_3(\a,\b,\half(\c+\c'),\half(\c+\c'-1);\c,\c',\c+\c'-1)\label{eq:eq371}\\
F(a,b;a+b+\half;z)&=&F(2a,2b;a+b+\half;\half-\half(1-z)^\half).
\label{eq:eq372}\ena
(\ref{eq:eq371}) is written in ref.\cite{Burchnall,SM} and (\ref{eq:eq372}) is found in ref.\cite{HTF}.

Having discussed these interesting identities about the underlying $K3$ surface, let us analyze the expansion around $c=1$. For \mone,{} this singularity is known to be the point of enhanced gauge symmetry for $SU(2)\times SU(2)$\cite{KV}. Noticing that the two type of solutions (\ref{eq:eq331}) and (\ref{eq:eq332}) can be written as
\bea
&\sum_{n_1,n_2}&F(n_1+2n_2+{\lambda\over2},n_1+2n_2+{1-\lambda\over2};2n_2+1;({2\over a})^2c) \nonumber\\
&{}&\times{\G(n_1+2n_2+{\lambda\over2})\G(n_1+2n_2+{1-\lambda\over2}) \over \G(2n_2+1)\G(n_1+\half)\G(n_1+1)\G(n_2+1)^2}({2\over a}-1)^{2n_1}[({2\over a})^4{b\over4}c^2]^{n_2},\label{eq:eq381}\\
&\sum_{n_1,n_2}&F(n_1+2n_2+{\lambda\over2}+\half,n_1+2n_2+{1-\lambda\over2}+\half ;2n_2+1;({2\over a})^2c) \nonumber\\
&{}&\times{\G(n_1+2n_2+{\lambda\over2}+\half)\G(n_1+2n_2+{1-\lambda\over2}+\half) \over \G(2n_2+1)\G(n_1+{3\over2})\G(n_1+1)\G(n_2+1)^2}({2\over a}-1)^{2n_1+1}[({2\over a})^4{b\over4}c^2]^{n_2},
\label{eq:eq382}\ena
we will apply the ordinary formula of the analytic continuation of the hypergeometric functions\cite{HTF}:
\bea
&{}&F(a,b;c;z) = {\G(c)\G(c-a-b)\over\G(c-a)\G(c-b)}F(a,b;a+b-c+1;1-z)\nonumber\\
&{}&\qquad+{\G(c)\G(a+b-c)\over\G(a)\G(b)}(1-z)^{c-a-b}F(c-a,c-b;c-a-b+1;1-z).
\label{eq:eq39}\ena
For the logarithmic solutions with respect to the summation over $n_3$, we can use the corresponding formula listed in \cite{HTF}. In this way, we have the following list of the hypergeometric series around the point $a=2,b=0,c=1$:
\bea
&{}&\sum_{n_1,n_2,n_3}{\G(n_1+{\lambda\over2})\G(n_1+{1-\lambda\over2})\G(2n_1+2n_2-n_3-\half)\over \G(n_1-n_3+{\lambda\over2})\G(n_1-n_3+{1-\lambda\over2})\G(n_1+\half)\G(n_1+1)\G(n_2+1)^2\G(n_3+1)}\nonumber\\
&{}&\qquad\qquad\qquad\qquad\times({x_1\over4})^{n_1}({y_1\over4})^{n_2}(-z_1)^{n_3+\half},\label{eq:3sol1}\\
&{}&\sum_{n_1,n_2,n_3}{\G(n_1+{\lambda\over2}+\half)\G(n_1+{1-\lambda\over2}+\half)\G(2n_1+2n_2-n_3+\half)\over \G(n_1-n_3+{\lambda\over2}+\half)\G(n_1-n_3+{1-\lambda\over2}+\half)\G(n_1+{3\over2})\G(n_1+1)\G(n_2+1)^2\G(n_3+1)}\nonumber\\
&{}&\qquad\qquad\qquad\qquad\times({x_1\over4})^{n_1+\half}({y_1\over4})^{n_2}(-z_1)^{n_3+\half},\label{eq:3sol2}\\
&{}&\sum_{n_1,n_2,n_3}{\G(n_1+2n_2+n_3+{\lambda\over2})\G(n_1+2n_2+n_3+{1-\lambda\over2})\G(n_1+{\lambda\over2})\G(n_1+{1-\lambda\over2}) \over \G(2n_1+2n_2+n_3 + \half)\G(n_1+\half)\G(n_1+1)\G(n_2+1)^2\G(n_3+1)}\nonumber\\
&{}&\qquad\qquad\qquad\qquad \times({x_1z_1^2\over4})^{n_1}({y_1z_1^2 \over4})^{n_2}z_1^{n_3},\label{eq:3sol3}\\
&{}&\sum_{n_1,n_2,n_3}{\G(n_1+2n_2+n_3+{\lambda\over2}+\half)\G(n_1+2n_2+n_3+{1-\lambda\over2}+\half)\G(n_1+{\lambda\over2}+\half)\G(n_1+{1-\lambda\over2}+\half)\over \G(2n_1+2n_2+n_3 + {3\over2})\G(n_1+{3\over2})\G(n_1+1)\G(n_2+1)^2\G(n_3+1)}\nonumber\\
&{}&\qquad\qquad\qquad\qquad \times({x_1z_1^2\over4})^{n_1+\half}({y_1z_1^2 \over4})^{n_2}z_1^{n_3},\label{eq:3sol4}
\ena
where the variables around the singular point is taken to be
\bea
x_1=4({{2\over a}-1 \over 1-({2\over a})^2c})^2,\quad y_1 = {({2\over a})^4bc^2 \over  (1-({2\over a})^2c)^2}, \quad z_1 = 1-({2\over a})^2c.
\label{eq:variable}\ena
We can also obtain the  logarithmic solutions with respect to the index $n_2$  either by shifting the indices and taking derivative or by using Barnes type representation as has been discussed before. In total, we have 8 independent solutions. The cycle of the models around this conifold locus are to be represented by using these solutions, which we will not attempt to analyze here.  

Around $z_1 =0$, the solutions (\ref{eq:3sol1}),(\ref{eq:3sol2}) and their logarithmic solutions reduce to
the functions:
\bea
F_4(-\fourth, \fourth,\half,1;x_1,y_1),\quad x_1^\half F_4(\fourth, {3\over4},{3\over2},1;x_1,y_1),
\ena
and their logarithmic solutions. Note that these does not depend on $\lambda$. In other words, the leading contribution is universal to the models. 

It turns out that these Appell functions can be written in terms of two type of hypergeometric functions. In order to see this, we use the following formula:
\bea
&{}&F_4(\a,\a+\half,\half,\c,x,y)\nonumber\\
&{}&\qquad\qquad\qquad\qquad=\half(1+x^\half)^{-2\a}F(\a,\a+\half;\c:{y\over(1+x^\half)^2})\nonumber\\
&{}&\qquad\qquad\qquad\qquad+\half(1-x^\half)^{-2\a}F(\a,\a+\half;\c:{y\over(1-x^\half)^2})\label{eq:3id1}\\
&{}&F_4(\a+\half,\a+1,{3\over2},\c,x,y)\nonumber\\
&{}&\qquad\qquad\qquad\qquad={-1\over 4\a x^\half}(1+x^\half)^{-2\a}F(\a,\a+\half;\c:{y\over(1+x^\half)^2})\nonumber\\
&{}&\qquad\qquad\qquad\qquad+{1\over 4\a x^\half}(1-x^\half)^{-2\a}F(\a,\a+\half;\c:{y\over(1-x^\half)^2})\label{eq:3id2}
\ena
The first formula (\ref{eq:3id1}) has been obtained in Ref\cite{Appell} and the second formula (\ref{eq:3id2}) can be obtained quite similarly. These results are derived as consequences of the following simple identities about hypergeometric series of the form\cite{HTF}:
\bea
F(a,a+\half;\half;z)&=& \half[(1+z^\half)^{-2a}+(1-z^\half)^{-2a}],\nonumber\\
F(a+\half,a+1;{3\over2};z)&=& {1\over 4az^\half}[-(1+z^\half)^{-2a}+(1-z^\half)^{-2a}].
\ena
 It is also easy to get the formula for logarithmic solutions. 
By using these identities for $\a=-\fourth$, we find that these are precisely the functions describing the solutions of the Seiberg-Witten's $SU(2)$ Yang-Mills theories\cite{SW,KLT}. Therefore, we find that this conifold corresponds to the point of enhanced gauge symmetry for the gauge groups $SU(2)\times SU(2)$, which is universal to these three moduli models. The correspondence to the S-W curve for $SU(2)\times SU(2)$ is given by
\bea
a = 2 + {\e \over2}(u_1-u_2),\quad b=\e^2\Lambda^4,\quad c= 1-\e u_2,
\ena
where $u_1$ and $u_2$ are the moduli of the rigid theory for $SU(2)\times SU(2)$.

\sect{Expansion around $(a=0, b=0, c=1)$}

We have seen that the conifold point for the enhanced $SU(2)\times SU(2)$ gauge symmetry is universal to the models. Let us next consider another singular point $(a=0, b=0, c=1)$, which is known to be the point for enhanced $SU(3)$ gauge symmetry for the model \mone.\cite{KV} By applying  the analytic continuation:
\bea
F(a,b;c;z) &=& {\G(c)\G(b-a)\over\G(b)\G(c-a)}(-z)^{-a}F(a,1-c+a;1-b+a;z^{-1})\nonumber\\
&+& {\G(c)\G(a-b)\over\G(a)\G(c-b)}(-z)^{-b}F(b,1-c+b;1-a+b;z^{-1}),
\label{eq:3ana1}\ena
to the expression (\ref{eq:eq31}) and  a quadratic transformation:
\bea
F(a,b;2b;z) = (1-\half a)^{-a}F(\half a,\half a+\half;b+\half;({z\over 2-z})^2), 
\ena
or just applying the formula (\ref{eq:3ana1}) to the expression (\ref{eq:eq32}), we see that the solutions are mapped to the following type of solutions:
\bea
&{}&\sum_{n_1,n_2,n_3}{\G(n_1+2n_2+n_3 +{\lambda \over2})\G(n_1+2n_2+n_3+{\lambda \over2}+\half)\over \G(n_1+\lambda+\half)\G(n_1+1)\G(n_2+1)^2\G(n_3+2n_2+1)\G(n_3+1)}\nonumber\\
&{}&\qquad\qquad\qquad\qquad\times({\half a \over 1-\half a})^{2n_1+\lambda}({b\over 4})^{n_2}({c \over (1-\half a)^2})^{n_3+2n_2},\label{eq:41}\\
&{}&\sum_{n_1,n_2,n_3}{\G(n_1+2n_2+n_3 +{1-\lambda \over2})\G(n_1+2n_2+n_3+{1-\lambda \over2}+\half)\over \G(n_1+1-\lambda+\half)\G(n_1+1)\G(n_2+1)^2\G(n_3+2n_2+1)\G(n_3+1)}\nonumber\\
&{}&\qquad\qquad\qquad\qquad\times({\half a \over 1-\half a})^{2n_1+1-\lambda}({b\over 4})^{n_2}({c \over (1-\half a)^2})^{n_3+2n_2},\label{eq:42}
\ena
and the logarithmic solutions with respect to the indices $n_2$ and $n_3$. Note that for \mfour $(\lambda =\half)$, these solutions degenerate and the logarithmic solution will appears by the analytic continuation. These difference can be understood in the framework of the underlying torus in the following way.  When we set $b=c=0$, the periods reduces to those of torus. The torus corresponding to models \mone, \mtwo{} and \mthree{} are the elliptic singular curves and $a=0$ corresponds to the critical point where there is no logarithmic behavior, whereas $P^{1,1,1,1}[2,2]$ does not describe singular curve. In order to understand  the curve, it may be better to use a formal analogy given in ref.\cite{MS2}. That is, by choosing derivation from the critical points, $N=2$ $SU(2)$ Yang-Mills theories with $N_f =1,2,3$ can be identified by the elliptic singular curves described above. On the other hand, the theory with $N_f =0$ can be identified with $P^{1,1,1,1}[2,2]$ whose moduli!
 are identifies with the deviati

on from the dyon or monopole points\cite{MS2}. Therefore, it is natural to find logarithmic solutions around these points. Keeping in mind the difference, we are going to make another transformation.\\
The solution (\ref{eq:41}) can be written in the form:
\bea
&{}&\sum_{n_1,n_2}F(n_1+2n_2+{\lambda \over2},n_1+2n_2+{\lambda\over2}+\half;2n_2+1;{c \over (1-\half a)^2})\nonumber\\
&{}&{\G(n_1+2n_2+{\lambda \over2})\G(n_1+2n_2+{\lambda\over 2}+\half)\over \G(2n_2+1)\G(n_1+\lambda+\half)\G(n_1+1)\G(n_2+1)^2} ({\half a \over 1-\half a})^{2n_1+\lambda}({b\over 4})^{n_2}({c \over (1-\half a)^2})^{2n_2}.{}
\label{eq:34}\ena
We next apply the formula (\ref{eq:3ana1}) to get
\bea
&\sum_{n_1,n_2}&F(n_1+2n_2+{\lambda\over2},n_1+2n_2+{\lambda\over2}+\half;2n_1+2n_2+\lambda +\half;1-{c\over(1-\half a)^2})\nonumber\\
&\times&{\G(\half -\lambda -2n_1-2n_2)\G(n_1+2n_2+{\lambda\over2})\G(n_1+2n_2+{\lambda\over2}+\half) \over \G(1-{\lambda\over2}-n_1)\G(\half-{\lambda\over2}-n_1)\G(n_1+\lambda+\half)\G(n_1+1)\G(n_2+1)^2}\nonumber\\
&{}&\qquad\qquad\times({\half a \over 1-\half a})^{2n_1+\lambda}({b\over 4})^{n_2}({c \over (1-\half a)^2})^{2n_2},\label{eq:351}\\
&\sum_{n_1,n_2}&F(-n_1+1-{\lambda\over2},-n_1+\half-{\lambda\over2};{3\over2}-\lambda -2n_1-2n_2;1-{c\over(1-\half a)^2})\nonumber\\
&\times&{\G(2n_1+2n_2+\lambda-\half)\G(n_1+2n_2+{\lambda\over2})\G(n_1+2n_2+{\lambda\over2}+\half) \over \G(n_1+2n_2+{\lambda\over2})\G(n_1+2n_2+{\lambda\over2}+\half)\G(n_1+\lambda+\half)\G(n_1+1)\G(n_2+1)^2}\nonumber\\
&{}&\qquad\qquad\times(1-{c\over(1-\half a)^2})^{\half-\lambda -2n_1-2n_2}({\half a \over 1-\half a})^{2n_1+\lambda}({b\over 4})^{n_2}({c \over (1-\half a)^2})^{2n_2}.\label{eq:352}
\ena
In order to simplify the coefficient and the variables, we apply a quadratic transformation:
\bea
F(a,a+\half;c;z)=(1-z)^{-a}F(2a,2c-2a-1;c;\half -\half(1-z)^{-\half}),
\label{eq:36}\ena
to (\ref{eq:351}). For the solution (\ref{eq:352}), we apply a quadratic transformation;
\bea
&{}&F(a,a+\half;c;z)\nonumber\\
&{}&\qquad = (1-z)^{-a}(\half+\half(1-z)^{-\half})^{-c+1}F(c-2a,2a-c+1;\half -\half(1-z)^{-\half}), 
\ena
which follows from the formula (\ref{eq:36}) and 
\bea
F(a,b;c;z) = (1-z)^{c-a-b}F(c-a,c-b,c;z).
\ena
The solution (\ref{eq:42}) and the logarithmic solutions with respect to the indices $n_3$ can be transformed quite similarly. In this way, we have solutions around the conifold singularity in the form:
\bea
&\sum_{n_1,n_2,n_3}&{\G(2n_1+2n_2-n_3+\lambda+\half)\G(2n_2+n_3+\half)({x\over4})^{n_1+{\lambda\over2}}({y\over4})^{n_2}z^{n_3+\half} \over \G(n_1+\lambda+\half)\G(n_1+1)\G(n_2+1)^2\G(n_3+1)\G(2n_2-n_3+\half)}\label{eq:4sol1} \\
&\sum_{n_1,n_2,n_3}&{\G(2n_1+2n_2-n_3+1-\lambda+\half)\G(2n_2+n_3+\half)({x\over4})^{n_1+{1-\lambda\over2}}({y\over4})^{n_2}z^{n_3+\half} \over \G(n_1+1-\lambda+\half)\G(n_1+1)\G(n_2+1)^2\G(n_3+1)\G(2n_2-n_3+\half)}\label{eq:4sol2} \\
&\sum_{n_1,n_2,n_3}&{\G(2n_1+4n_2+n_3+\lambda)\G(2n_1+n_3+\lambda)\over \G(n_1+\lambda +\half)\G(n_1+1)\G(n_2+1)^2\G(n_3+1)\G(2n_1+2n_2+n_3+\lambda+\half)} \nonumber \\
&{}&\qquad\qquad\qquad\qquad\qquad\times({xz^2\over4})^{n_1+{\lambda\over2}}({yz^2\over4})^{n_2}z^{n_3},\label{eq:4sol3}\\
&\sum_{n_1,n_2,n_3}&{\G(2n_1+4n_2+n_3+1-\lambda)\G(2n_1+n_3+1-\lambda)\over \G(n_1+1-\lambda +\half)\G(n_1+1)\G(n_2+1)^2\G(n_3+1)\G(2n_1+2n_2+n_3+1-\lambda+\half)}\nonumber\\
&{}&\qquad\qquad\qquad\qquad\qquad\times({xz^2\over4})^{n_1+{1-\lambda\over2}}({yz^2\over4})^{n_2}z^{n_3},\label{eq:4sol4}
\ena
where the variables $x,y$ and $z$ are given by
\bea
x= {a^2 \over 4c(1-{1-\half a\over \sqrt{c}})^2}, y={b \over 4(1-{1-\half a\over \sqrt{c}})^2}, z = \half(1-{1-\half a\over \sqrt{c}}).
\ena
There are four other solutions logarithmic with respect to the indices $n_2$, which we can obtain  by shifting $n_2$ and differentiating with respect to the deviations or by using the Barnes type-representations as before. For the model \mfour,  these solutions should be regularized by using Barnes type representation and other logarithmic solutions are also required. We are not attempt to list the solutions because these may not be suffered from essential difficulties.
 
Around the point $z=0$, the solutions (\ref{eq:4sol1}) and (\ref{eq:4sol2}) and their logarithmic solutions reduce to the Appell functions
\bea
&{}&x^{\lambda\over2}F_4({\lambda\over2}-\fourth, {\lambda\over2}+\fourth,\lambda+\half,1;x,y),\nonumber\\
&{}&x^{1-\lambda\over2}F_4({1-\lambda\over2}-\fourth, {1-\lambda\over2}+\fourth,1-\lambda+\half,1;x,y),
\ena
and their logarithmic solutions.
In other branch, these are written by functions\cite{HTF}:
\bea
&{}&x^\fourth F_4({\lambda\over2}-\fourth,-{\lambda\over2}+\fourth,\half;1;{1\over x},{y \over x}),\nonumber\\
&{}&x^{-\fourth}F_4({\lambda\over2}+\fourth,-{\lambda\over2}+{3\over4},{3\over2},1;{1\over x},{y \over x}),
\label{eq:f4}\ena
and the corresponding logarithmic functions. 

It is interesting that the leading behavior of these models can be described by means of Appell functions. 
Contrary to the case of the conifold point $(a=2,b=0,c=1)$, however, the parameters of the Appell functions depend on the models, which means that the underlying Riemannian surfaces are not universal.  In the next section, we are going to analyze the connection to the Riemannian surfaces.
\ 
\sect{Underlying Riemannian Surfaces and Appell functions}
In the previous section, we have seen that the expansion around the conifold singularity can be described in the form of the hypergeometric series and that the leading behavior can be written in terms of Appell function of type $F_4$. However, the underlying Riemannian surface depends on models. A convenient approach for discussing the relation to the Riemannian surface are given in ref.\cite{KLMVW}. We just apply the method to the models that we are considering here.\\
\noindent\underline{\mone}{}$(\lambda={1\over6})$:

This models has been investigated in ref.\cite{KLMVW} so that we will repeat their argument for completeness.

By introducing the variables $z_4 = z_0^\half \zeta^{1\over 24}, z_5 = z_0^\half b^{1\over24}\zeta^{-{1\over24}}$, the defining polynomial of the model can be rewritten as\cite{KLMVW}
\bea
f = {1 \over 24}(\zeta+{b\over\zeta}+2)z_0^{12}+\half z_1^2&+&\third z_2^3+{1\over12}z_3^{12}\nonumber\\
&{}&+{1\over 6\sqrt{c}}(z_0z_3)^6-({a\over\sqrt{c}})^{1\over6}z_0z_1z_2z_3.
\ena
Around the singular point $z_0=1, z_3= (-1)^{1\over6}$, by scaling the moduli as
\bea
a=-\epsilon {2\over 3\sqrt{3}}u^{3\over2}, b= \e^2\Lambda^6, 1-a-c = \e v,
\ena
we can expand the polynomial in the form:
\bea
f = {1 \over 24} \e [z + {\Lambda^6\over z}+ 2(z^3-uz-v)+ \rho^2 + \sigma^2] + O(\e^2),
\ena
which corresponds to the curve given for N=2 super-Yang-Mills theory of the group $SU(3)$\cite{KLT}.
It is known that the solution of the Picard-Fuchs equation can be written in terms of Appell functions with parameters given in (\ref{eq:f4}) for $\lambda = {1\over6}$. 

\noindent\underline{\mtwo}{}$(\lambda=\third)$:

The defining polynomial of this model can be written as
\bea
f= {1\over16}(\zeta+{b\over\zeta}+2)z_0^8+\half z_1^2&+&\fourth z_2^4+{1\over8}z_3^8\nonumber\\
&{}&+{1 \over 4\sqrt{c}}(z_0z_3)^4-({a\over\sqrt{c}})^\fourth z_0z_1z_2z_3,
\ena
where we have set $z_4 = z_0^\half \zeta^{1\over16}, z_5=z_0^\half b^{1\over16}\zeta^{-{1\over16}}.$\\
By writing the moduli around singular point as
\bea
a = \e {u^2 \over2}, b=\e^2 \Lambda^8, c= 1 - \e v^2,
\ena
the defining polynomial can be written as
\bea
f = {1 \over 16}\e[ z + {\Lambda^8\over z} + 2(x^4-2ux^2+v^2) + \rho^2 + \sigma^2] + O(\e^2),
\label{eq:mtwo}\ena
which corresponds to  the integralbe curve of the type $C_2$\cite{MW}.  The ordinary rigid curve of the gauge group $C_2$ are identified by the curve ${\hat C}_2$ which is dual to the curve (\ref{eq:mtwo})\cite{MW}.  Although this parameterization shows us  the curve of  $A$-type singularity , we do not have any rigid interpretation of the curve. It is interesting that the solutions of the above curve can be solved in terms of Appell functions by choosing the variables as
\bea
x = ({u^2 \over 2v^2-u^2})^2, \qquad y= {\Lambda^8 \over (v^2-{u^2\over2})^2},
\ena
which can be checked by explicit evaluation of the periods of the curve.\\
\noindent\underline{\mthree}{}$(\lambda=\fourth)$:

The defining polynomial can be written in the form:
\bea
f = {1\over 12}(\zeta+{b\over\zeta}+2)z_0^6+\third z_1^3&+& \third z_2^3+{1\over6}z_3^6\nonumber\\
&{}&+{1\over3\sqrt{c}}(z_0z_3)^3-({a\over\sqrt{c}})^\third z_0z_1z_2z_3,
\ena
where we have set $z_4=z_0^\half\zeta^{1\over12}, z_5=z_0^\half b^{1\over12}\zeta^{-{1\over12}}.$ By parameterizing $a,b$ and $c$ as
\bea
a = -{2 \over 3\sqrt{3}}u^{3\over2}\epsilon, b= \epsilon^2\Lambda^6, c=1+({2\over3\sqrt{3}}u^{3\over2}-v)\epsilon,
\ena
we can expand the defining polynomial in the form:
\bea
f={1\over12}\epsilon[z+{\Lambda^6\over z}+ 2(x^3-ux-v) + (x+{2\over \sqrt{3}}u^\half)\rho^2 +\sigma^2]+O(\e^2),
\label{eq:m3effective}\ena
which is one of the typical  $D$-type singular curve\cite{Lerche}. In the evaluation of the periods, the factor $ (x+{2\over \sqrt{3}}u^\half)$ in front of the variable $\rho$ can be absorbed in the re-defenition of $\rho$, leaving the factor $ (x+{2\over \sqrt{3}}u^\half)^{-\half}$ in the period integral. This factor can be absorbed in the re-definition of $x$ by $x= y^2 -{2\over\sqrt{3}}u^\half$. Therefore, the curve (\ref{eq:m3effective}) can be written as
\bea
W = z+{\Lambda^6 \over z}+2[(y^2-{2\over\sqrt{3}}u^\half)^3-u(y^2-{2\over\sqrt{3}}u^\half) -v] + \rho^2 + \sigma^2.
\ena
It is interesting that the period integral can be written in terms of Appell functions with respect to the variables $x= {27\over4}{u^3\over v^2}$ and $y={\Lambda^6 \over v^3}$.  This result can be checked by a direct evaluation of the periods of the Riemannian surface corresponding to the curve (\ref{eq:m3effective}). 

\noindent\underline{\mfour}{}$(\lambda=\half)$:

The above three models are within the class analyzed in Ref.\cite{KLMVW}, which can be written in the form of $A$-$D$-$E$ classification of the singular points\cite{KLMVW,Lerche}. However, this model is out of the framework of these classification.  

The defining polynomials of this model can be written by
\bea
f_1 &=& {1\over8}(\zeta+{b\over\zeta}+2)z_0^4 + \half z_1^2 + \fourth z_2^4 -z_5z_6 + {1\over2\sqrt{c}}(z_0z_2)^2\nonumber\\
f_2 &=& \half z_5^2 +\half z_6^2 - ({a\over\sqrt{c}})^\half z_0z_1z_2,
\ena
where we have set $z_3 = z_0^\half\zeta^{1\over8}, z_4 = z_0^\half b^{1\over8}\zeta^{-{1\over8}}$.

By eliminating  $z_1$ from these polynomials, we find that the polynomial can be expressed in the form:
\bea
f_1 = {1 \over8}\epsilon[ z+{\Lambda^4\over z} + 2(x^2-u) + (v^2-x^2)\rho^2 + \sigma^2] + O(\e^2),
\ena
 where the identification for the weak coupling limit is given by
$a = \epsilon {v^2 \over2}, b= \epsilon^2 \Lambda^4, c = 1 - \epsilon u$. Let us re-write the curve in such a way that correspondence to Riemannian surface will be clearer. As has been discussed for the model \mthree, $\rho$ can be re-scaled but instead, $x$ should be written by sine functions in this case. Namely. by suitable change of variables, the curve can be written in the form:
\bea
W &=& z+{\Lambda^4 \over z} + 2 v^2[\sin^2{ x \over v} - \sin^2{ e \over v}] + \rho^2 + \sigma^2\nonumber\\
&=& z+{\Lambda^4 \over z} + 2 v^2\sin{(x+e)\over v}\sin{(x-e)\over v} + \rho^2 + \sigma^2,
\ena
where we have defined $e$ as ${e\over v} = \arcsin{\sqrt{u}\over v}$.
which can be considered as  an elliptic  (periodic) extension of the $SU(2)$ Seiberg-Witten curve. Since sine functions can be written as $\sin\pi x = \pi x \prod_{n=1}^\infty (1-{x\over n})(1+{x\over n})$, the curve can be formally regarded as a periodically reduced $SU(\infty)$ supersymmetric Yang-Mills theory with a renormalized dynamical mass scale. In the limit $T= \infty$, the theory reduces to the ordinary $SU(2)$ Yang-Mills theory. It is interesting that the conifold singularities generate such integrable curve with infinite number of singular points.

\sect{Discussions}
We have shown that expansion around the conifold singularities can be obtained in the form of hypergeometric series in the three-moduli models constructed by sequence of fibrations. It was shown that the the structure of the singularities can be analyzed quite systematically. If we make identification of the cycles, we can  obtain prepotentials around the conifold singularities.  

The most important problem we have to analyze is whether we can obtain  these elegant expansion for generic models. Although the method used here rely heavily on the form of the hypergeometric series at large moduli, which may be related to the fact that these models are constructed by sequence of fibrations, the resulting expansion may hopefully have some characteristic property which can be generalized to all models. Clearly, the study is closely related to the classification of the conifold singularities, which may reviel un-known phase transition and may produce new kind of rigid interpretation.

We would like to thank members of Physics Department in Hokkaido University for discussions. We also thank S. Hosono for pointing out references.


\end{document}